\documentclass[a4paper,
               biblatex,     
               ]{jacow}
\usepackage{pdfpages ,multirow ,ragged2e}
\usepackage{caption}
\usepackage{subcaption}

\makeatletter%
	\ifboolexpr{bool{xetex}}
	 {\renewcommand{\Gin@extensions}{.pdf,%
	                    .png,.jpg,.bmp,.pict,.tif,.psd,.mac,.sga,.tga,.gif,%
	                    .eps,.ps,%
	                    }}{}
\makeatother

\ifboolexpr{bool{xetex} or bool{luatex}}
 {}                                      
 {\usepackage[utf8]{inputenc}} 

\usepackage[USenglish]{babel}

\ifboolexpr{bool{jacowbiblatex}}%
 {%
  \addbibresource{THP033.bib}
 }{}
\begin{document}

\title{Feasibility of proton bunch compression in an operational high-power accelerator\thanks{This manuscript has been authored by UT-Battelle, LLC, under Contract No. DE-AC05-00OR22725 with the U.S. Department of Energy. The United States Government retains, and the publisher, by accepting
the article for publication, acknowledges that the United States Government retains a non-exclusive, paid-up, irrevocable, world-wide license to publish or reproduce the published form of this manuscript, or allow others to do so, for United States Government purposes. The Department of Energy will provide public access to these results of federally sponsored research in accordance with the DOE Public Access Plan (http://energy.gov/downloads/doe-public-access-plan)}}

\author{
A.\,Hoover\thanks{hooveram@ornl.gov}, 
V.\,Morozov, A.\,Narayan, 
M.\,Piller \\
Oak Ridge National Laboratory, Oak Ridge, TN, USA
}
	
\maketitle

\begin{abstract}
A muon collider (MC) would require a high-power proton driver to generate intense muon beams at the start of the accelerator chain. Like other high-power facilities, the driver would accumulate intense proton bunches via charge-exchange injection from a linac into a ring. However, unlike other facilities, the MC would require an extreme longitudinal compression of the bunch. We aim to experimentally study the proposed compression scheme at the Spallation Neutron Source (SNS), which boasts a world-leading proton bunch intensity. In this paper, we describe the operating parameters of the SNS compared to the MC design, the capabilities of the current RF system, and the available beam diagnostics in the SNS accumulator ring. We also present initial simulations and experiments to judge the feasibility of the proposed research.
\end{abstract}

\section{Introduction}

A muon collider (MC) would require a high-power proton driver (PD) to generate intense muon beams at the start of the accelerator chain \cite{accettura_muon_2025}. Like other high-power facilities, the PD would accumulate intense proton bunches via charge-exchange injection from an H- ion linac into a ring. However, unlike other facilities, the MC would require an extreme longitudinal compression of the bunch to a few nanoseconds in length. Although the proposed compression scheme is straightforward, the envisioned beam power (2-4 MW), energy (5-10 GeV), repetition rate (10 Hz), and compressed bunch length (2 ns) would lead to an unprecedented charge density that could excite various resonances and instabilities, possibly leading to beam loss.

The Spallation Neutron Source (SNS) at Oak Ridge National Laboratory (ORNL) is a close analogue of the PD and could be used to experimentally test the bunch formation process at high space charge intensities \cite{morozov_ipac_2024}. This paper begins to investigate the feasibility of such experiments. In the following sections, we describe the SNS accelerator layout, operational parameters, and available beam diagnostics. We then discuss a set of initial simulations and experiments performed at the SNS in the past year. Finally, we mention several collective effects that could influence the beam dynamics during compression. 

\section{SNS facility and diagnostics}

\begin{figure}
     \centering
     \includegraphics[width=\columnwidth]{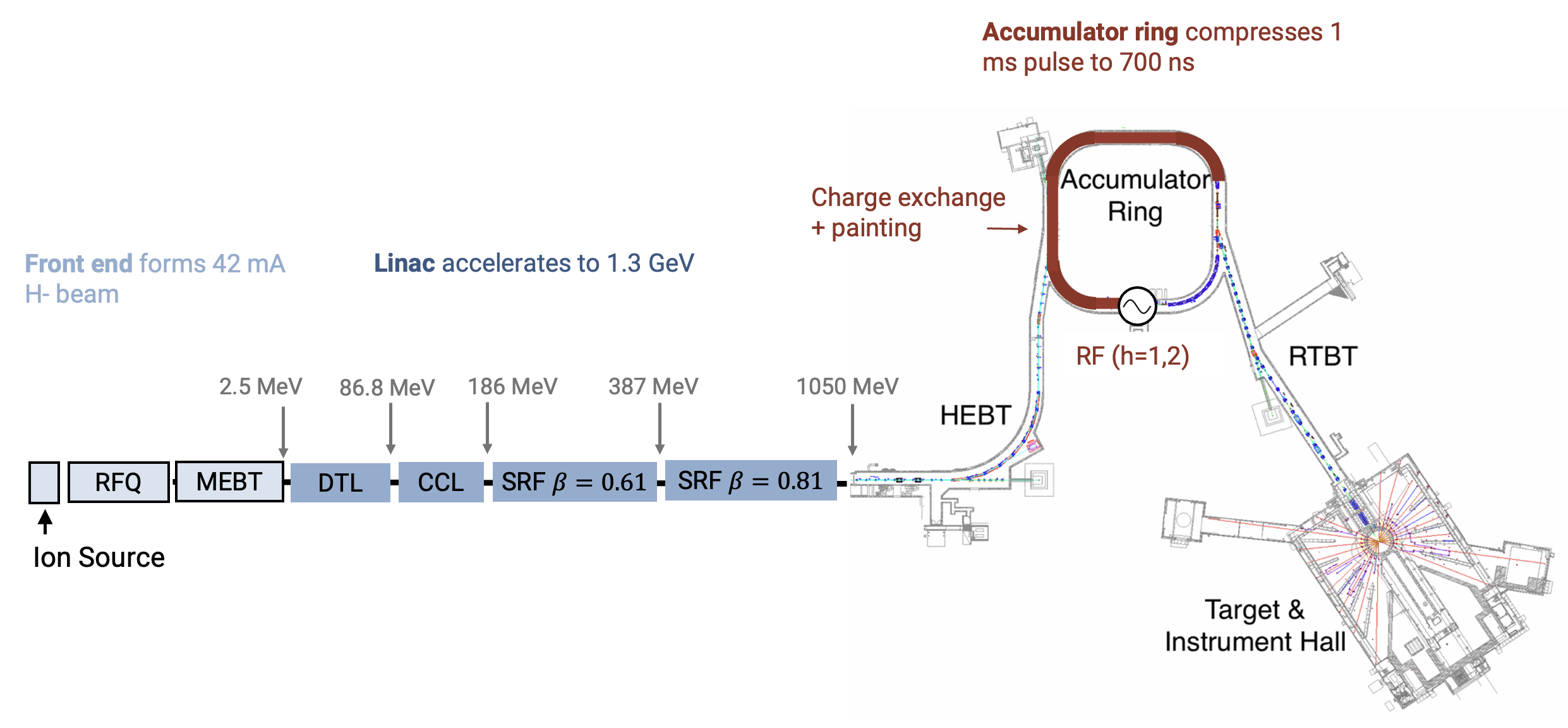}
     \caption{Overview of the SNS accelerator.}
     \label{fig:sns}
 \end{figure}

A diagram of the SNS accelerator is shown in Fig.~\ref{fig:sns}.
The SNS generates neutrons by colliding intense proton pulses with a liquid mercury target. Each so-called \textit{macropulse} on the target is the superposition of over 1000 \textit{minipulses} injected from a 400-meter linear accelerator (linac) into a 250-meter accumulator ring (AR) at 1.3 GeV energy. The accumulated beam is extracted and transported from the ring to the target in a 150-meter Ring-Target Beam Transport (RTBT). The SNS operates at a world-leading 1.8 MW beam power at 60 Hz repetition rate, with plans ramp the power from 1.8 to 2.8 MW over the next few years \cite{champion_ipac_2024}.

\begin{figure}
    \centering
    \includegraphics[width=0.90\columnwidth]{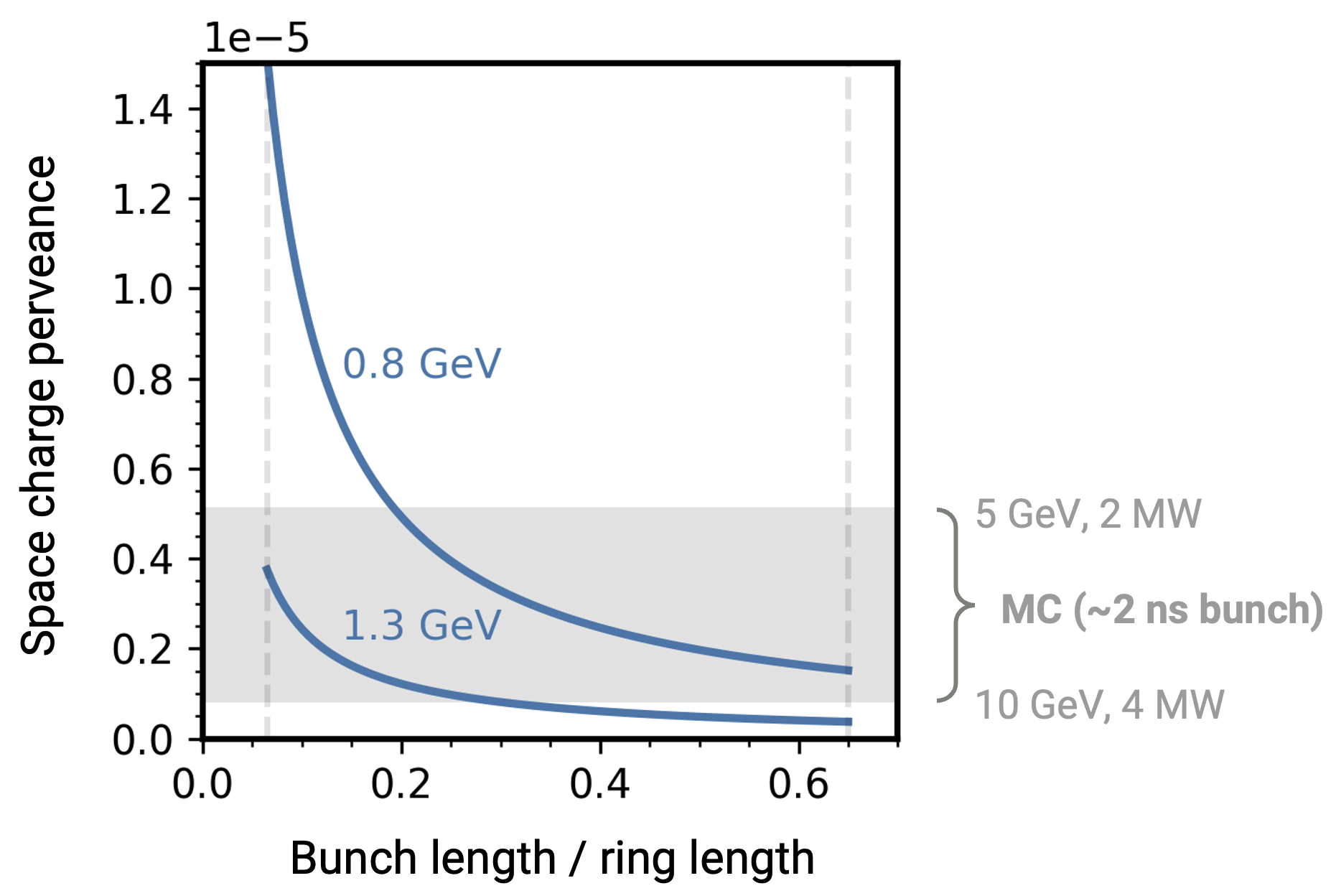}
    \caption{Space charge perveance vs. bunch length in the SNS at 2.8 MW beam power. The grey region is an estimated range for the MC proton driver at peak bunch compression.}
    \label{fig:perveance}
\end{figure}

Figure~\ref{fig:perveance} shows the space charge perveance $Q \propto \lambda / \beta^2 \gamma^3$ (where $\lambda$ is the longitudinal charge density and $\beta$, $\gamma$ are the relativistic factors) as a function of the beam energy at 2.8 MW beam power. The dashed line on the far right shows the nominal bunch length in the ring, while the line on the left shows an estimated minimum bunch length after compression. At the nominal beam energy of 1.3 GeV, strong compression would begin to probe the MC space charge regime. The SNS can also lower the beam energy to 0.8\,GeV by shutting off cavities at the end of the linac. At this lower energy, moderate to strong compression could probe the full range of possible space charge intensities in the MC.

A suite of diagnostics is available to measure the beam in the SNS accumulator ring (AR). In addition to beam position monitors (BPMs) and beam loss monitors (BLMs), an electron scanner \cite{blokland_recent_2013} provides instant turn-by-turn beam profiles along the $x$ and $y$ axes. The AR also hosts a beam instability monitor (BIM) \cite{evans_damping_2018} to measure fast-rising beam instabilities and a beam current monitor (BCM) to measure the longitudinal charge distribution on each turn. After the ring, four wire scanners (WS) provide slow but stable measurements of the one-dimensional transverse charge distribution along the horizontal, vertical, and diagonal axes at different locations in the RTBT. These wire scanners constrain both the two- and four-dimensional phase space distribution \cite{hoover_four_2024}, which can be estimated with tomographic methods. These diagnostics could provide a detailed, time-dependent view of the bunch during compression.

\section{Bunch compression methods}

The PD will utilize a quarter rotation in longitudinal phase space to compress the accumulated proton bunch. For a small-amplitude particle in a harmonic radiofrequency (RF) cavity, the synchrotron tune $\nu_z$ is given by Ref.~\cite{lee_accelerator_2018}
\begin{equation}
    \nu_z = \sqrt{\frac{ h q V |\eta_0| }{ 2 \pi \beta^2 E }},
\end{equation}
where $q$ is the particle charge, $E$ is the kinetic energy, $\beta$ is the velocity relative to the speed of light, $V$ is the RF voltage, $h$ is the RF harmonic number, and $\eta_0$ is the first-order slip factor linearly relating fractional changes in the revolution frequency $f$ to changes in longitudinal momentum $p$. High-power cavities and specially designed transverse optics will be used to maximize the synchrotron tune and compress the bunch in as few as tens of turns \cite{accettura_muon_2025}. The compression factor is determined by the initial distribution, which ideally has minimal energy spread and is confined to the linear portion of the RF bucket. The accumulator ring will be designed to optimize these initial beam parameters.

Let us now examine whether the proposed bunch compression scheme could be tested in the SNS. The SNS employs dual-harmonic ($h \in \{1, 2\}$) RF focusing in the ring. Two out of four available cavities are used during production, each providing voltages of 6-7 kV. If both $h=1$ cavities were set to their maximum voltage of 10 kV, a quarter synchrotron oscillation would be possible in approximately 400-600 turns, depending on the beam energy. It may be possible to increase the rotation speed by running all four cavities at the same $h=1$ harmonic. Thus, compression is possible, but it will be much slower than in the MC.

The maximum compression factor in the SNS will depend on the initial (accumulated) beam distribution. A small initial energy spread can be maintained during accumulation using \textit{barrier} (square well) RF waveforms \cite{holmes_barrier_2006}. There may be room to install a barrier cavity in the SNS ring, but this is an expensive option. Initial studies will seek to modify the existing RF system to approximate a barrier waveform. The following three ideas have been considered.

One idea is to turn off the RF focusing in the ring during accumulation. The eventual filamentation of the beam throughout the ring would create extraction losses, but such losses can be tolerated at the low repetition rates (1 Hz) typical of accelerator physics experiments. We found that operating with the cavities on, but with zero drive voltage, leads to significant beam-induced voltage in the cavity that exceeds the maximum cavity voltage well before full intensity. An alternative approach would be to short the RF gaps during accumulation and then unshort them immediately after accumulation, but this is not possible with the mechanical relay used in the present system.

\begin{figure}[t]
    \centering
    \includegraphics[width=1.0\columnwidth]{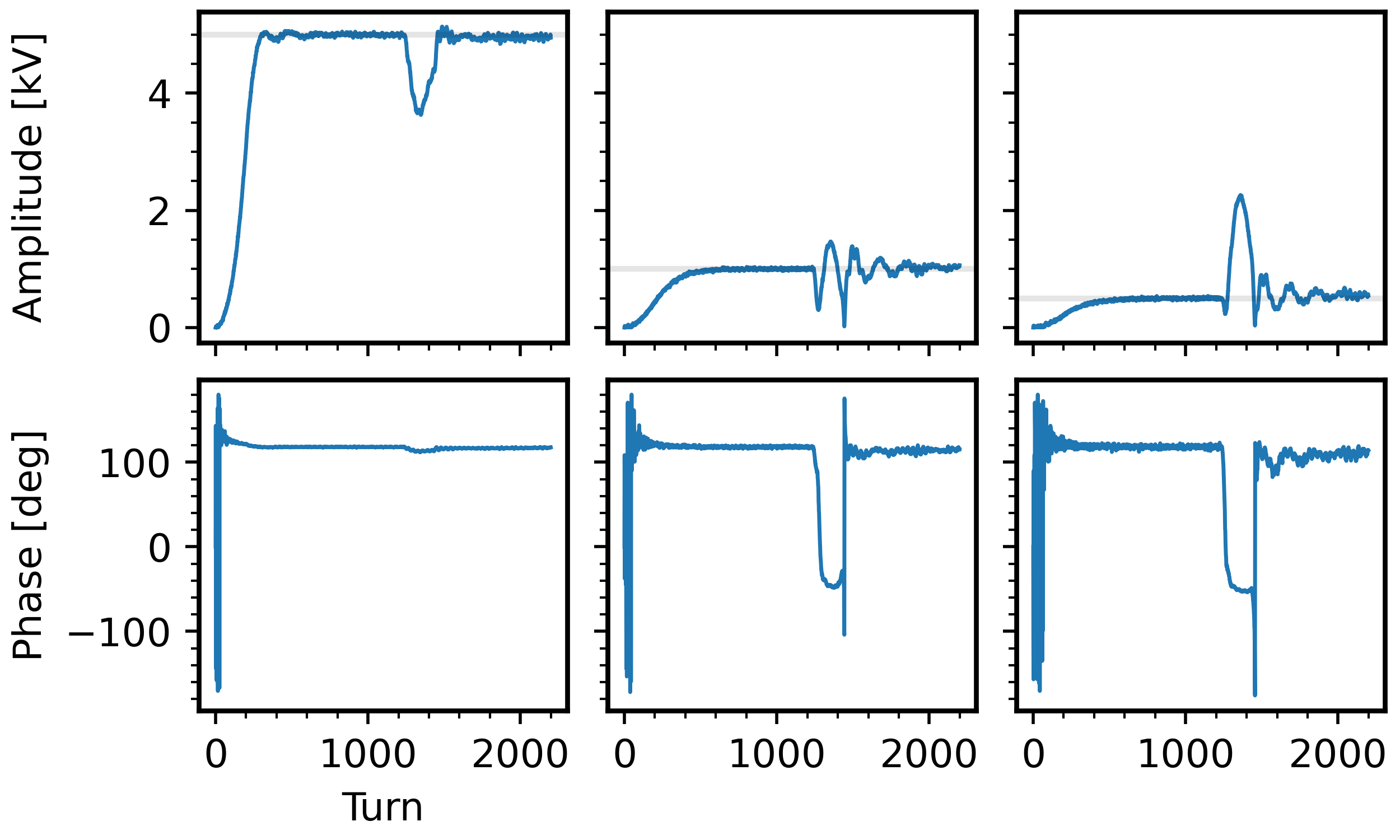}
    \caption{Measured RF amplitude and phase for various setpoints with feedback system engaged. Amplitude setpoints are plotted as faint grey horizontal lines (4 kV, 1 kV, 0.5 kV).}
    \label{fig:feedback}
\end{figure}

A second idea is to use the RF feedback control system to force the \textit{net} voltage (applied + beam-induced) to zero. The feedback system varies the RF drive amplitude and phase until the measured voltages and phases return to their setpoints. Figure~\ref{fig:feedback} shows the measured voltages and phases for three different setpoints, with the voltage decreasing from left to right, for 100 injected turns.
We find that the feedback system can correct for the beam-induced voltage after a large spike at the start of injection. Future research will aim to damp this voltage spike.

Third, we may approximate barrier waveforms using additional cavity harmonics. Four harmonics can generate a fairly flat waveform in the center of the bunch. This approach was studied in Ref.~\cite{holmes_barrier_2006}. The RF cavities in the SNS are designed for $h = 1$ or $h = 2$ operation, but could potentially be modified to run at $h = 3$ or $h = 4$ with the installation of new ferrite rings. Future work will use beam physics simulations to study this scheme in more detail.


Two minor issues are also worth mentioning. First, the SNS has just one RF system which must be used for both accumulation and compression. The waveforms must therefore switch immediately between the accumulation and compression phases. A custom waveform generator has recently been developed to perform this task. Second, the beam storage time is limited by the injection kicker magnet waveforms to avoid overheating of the magnets. Extension of the storage time (after the 1000-turn accumulation) to 1000-2000 turns appears to be possible but has not been tested.

\section{Initial experiment and simulation benchmark}

\begin{figure}[t]
     \centering
     \begin{subfigure}[t]{0.342\columnwidth}
         \centering
         \includegraphics[width=\columnwidth]{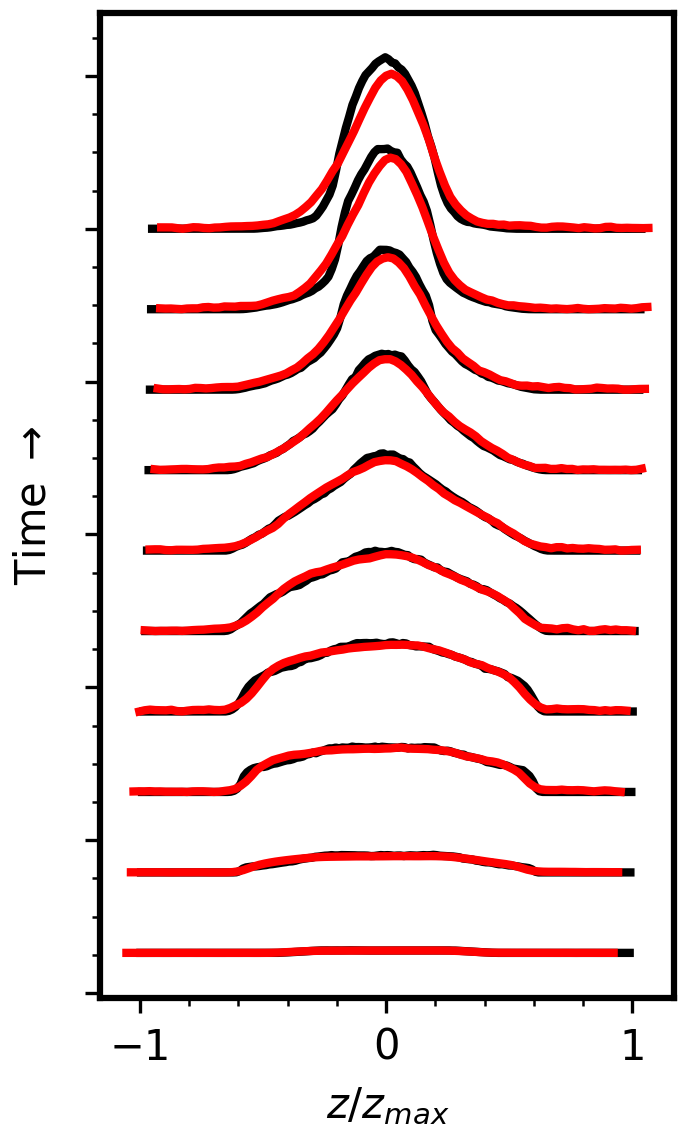}
         \caption{}
         \label{fig:exp_a}
     \end{subfigure}
     \hfill
     \begin{subfigure}[t]{0.57\columnwidth}
         \centering
         \includegraphics[width=\columnwidth]{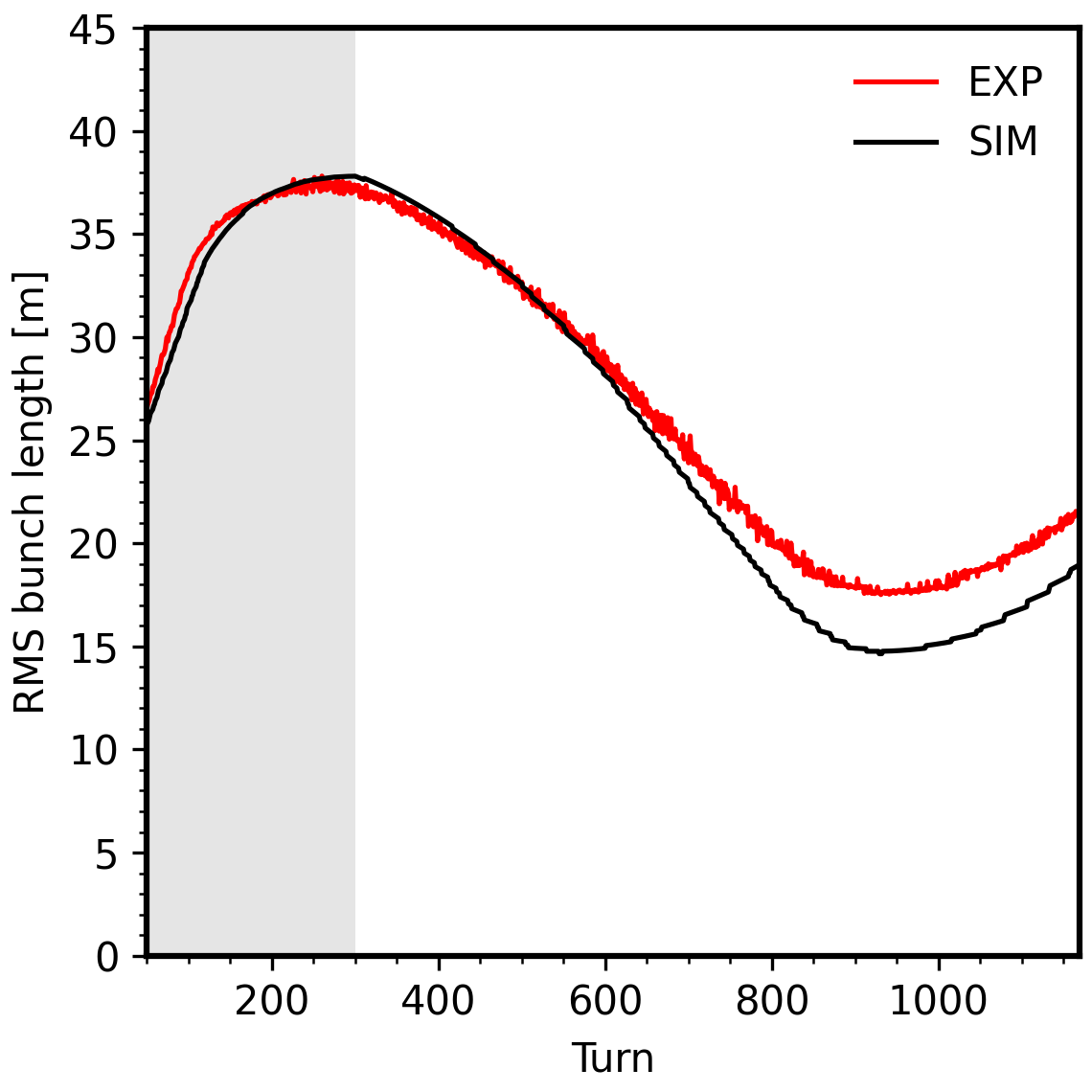}
         \caption{}
         \label{fig:exp_b}
    \end{subfigure}
    \caption{Simulation benchmark of low-intensity injection into fixed $h=1$ waveform. (a) BCM profiles; (b) rms bunch length (10\% threshold). The beam is accumulated over the first 300 turns (grey region).}
    \label{fig:exp}
\end{figure}

We performed an initial experiment to moderately compress a low-intensity bunch for simulation benchmarking purposes. We accumulated beam for 300 turns with a fixed 10 kV RF voltage at $h = 1$ frequency, with all other cavities shorted and the feedback system activated. The accumulated bunch was stored for 900 turns. Since the synchrotron period was much longer than the injection time, we did not expect the longitudinal distribution to rotate much during injection, and thus expected moderate compression during beam storage. Figure~\ref{fig:exp} compares the measured BCM profiles to simulation, showing reasonable agreement. 
The simulation used an initial Gaussian energy distribution and uniform spatial distribution with rms sizes chosen to fit the data, using a one-dimensional space charge solver for the longitudinal dynamics. Future work will use tomographic methods to infer the complete longitudinal phase space distribution from the measured profiles \cite{kelliher_ipac_2025}.

\section{Collective effects}

\begin{figure}[t]
    \centering
    \includegraphics[width=1.0\columnwidth]{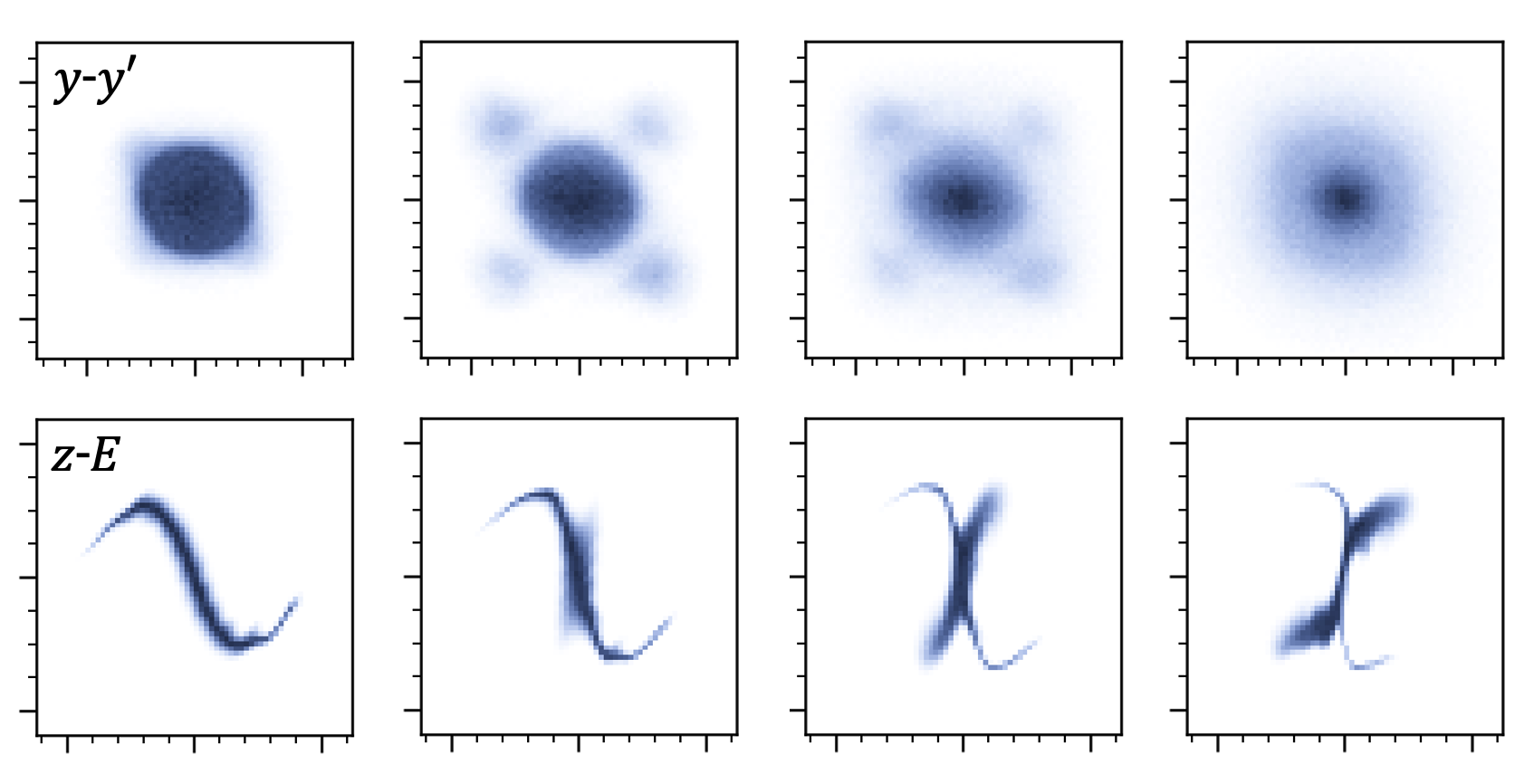}
    \caption{Simulated evolution in phase space during compression. Top: fourth-order resonance in transverse phase space. Bottom: characteristic \textit{x} shape in longitudinal phase space.}
    \label{fig:res}
\end{figure}

We hope to use bunch compression to study high-intensity beam physics in the SNS accumulator ring. Simulations predict a fourth-order resonance structure in the transverse phase space excited by the nonlinear space charge forces in the matched beam during compression, see Fig.~\ref{fig:res}.
The emergence of this resonance during compression was predicted by Hoffman \cite{hofmann_space_2017} and could, in principle, be measured in the SNS using tomographic methods. Simulations also predict strong nonlinear space charge effects in the longitudinal phase space. The competition between the nonlinear RF focusing and nonlinear space charge defocusing was briefly investigated in Ref.~\cite{ng_space_2001}; these studies should be continued and applied to SNS experiments. Especially important is the impact of space charge on the optimal RF drive amplitude and phase during compression. Finally, electron-proton (EP) instabilities have been observed in the SNS \cite{shishlo_electron_2006} but are damped by the RF focusing during production \cite{evans_damping_2018}. It is unknown how bunch compression will affect the EP instability. Future work will focus on modeling these collective effects in accelerator physics codes such as PyORBIT \cite{shishlo_electron_2006}, ImpactX \cite{mitchell_impactx_2023}, and Xsuite \cite{iadarola_xsuite_2023} in collaboration with other facilities \cite{simons_ipac_2024}.

\section{Conclusion}

With significant bunch compression and energy modulation, the SNS could study the physics of beams at very high space charge intensities. This experimental program would be relevant to the design of a future muon collider, but also of general interest to the high-intensity accelerator community. Initial studies suggest that bunch compression could be possible without major hardware upgrades, but additional experiments are needed. Data, simulations, and calculations presented in this paper can be found in an online repository \cite{hoover_zenodo_2025}.

\printbibliography

\end{document}